\documentclass[]{spie}
\usepackage{graphicx} 
\usepackage{xcolor}
\usepackage{amsmath,amsfonts,amssymb}
\usepackage{url}
\usepackage{siunitx}
\usepackage{listings}
\usepackage{subcaption}
\usepackage{xspace}  
\usepackage[colorlinks=true, allcolors=blue]{hyperref}
\usepackage{lineno}
\usepackage[square,numbers]{natbib}

\usepackage{tikz,xcolor,hyperref}
\definecolor{lime}{HTML}{A6CE39}
\DeclareRobustCommand{\orcidicon}{%
    \begin{tikzpicture}
    \draw[lime, fill=lime] (0,0) 
    circle [radius=0.16] 
    node[white] {{\fontfamily{qag}\selectfont \tiny ID}};
    \draw[white, fill=white] (-0.0625,0.095) 
    circle [radius=0.007];
    \end{tikzpicture}
    \hspace{-2mm}
}
\newcommand{\orcid}[1]{\href{https://orcid.org/#1}{\orcidicon}}

\title{In Situ Measurements of the Reflectances of the LSSTCam Optics and Assessing the Impact of Optical Ghosts}

\author[1, 2, 3]{Aashay~Pai \orcid{0009-0008-9641-6065}}

\author[2, 3, 4, 5]{Alex~Drlica-Wagner \orcid{0000-0001-8251-933X}}

\author[6]{Lee~S.~Kelvin \orcid{0000-0001-9395-4759}}

\author[7]{Joshua~E.~Meyers \orcid{0000-0002-2308-4230}}

\author[8]{Elana~K.~Urbach
\orcid{0000-0002-3205-2484}}

\author[7]{Fritz~Mueller
\orcid{0000-0002-7061-4644}}

\author[6]{Robert~H.~Lupton
\orcid{0000-0003-1666-0962}}

\affil[1]{\small Department of Physics, University of Chicago, Chicago, IL 60637, USA}
\affil[2]{{\small Kavli Institute for Cosmological Physics, University of Chicago, Chicago, IL 60637, USA}}
\affil[3]{\small NSF-Simons AI Institute for the Sky (SkAI), 172 E. Chestnut St., Chicago, IL 60611, USA}
\affil[4]{\small Fermi National Accelerator Laboratory, Batavia, IL 60510, USA}
\affil[5]{\small Department of Astronomy \& Astrophysics, University of Chicago, Chicago, IL 60637, USA}
\affil[6]{\small Department of Astrophysical Sciences, Princeton University, Princeton, NJ 08544, USA}
\affil[7]{\small SLAC National Accelerator Laboratory, Menlo Park, CA, United States}
\affil[8]{\small Department of Physics, Harvard University, 17 Oxford St., Cambridge MA 02138, USA}

\authorinfo{Further author information: Aashay Pai (pai@uchicago.edu)}

\begin{document} 
\maketitle

\begin{abstract}
Optical ghosts are image artifacts caused by successive reflections of light between optical surfaces such as lenses, filters, and detectors. 
These artifacts are unavoidable due to the nonzero reflectances of optical elements and are a major source of contamination for low-surface-brightness science. 
We use optical ray tracing simulations tuned to observations from LSST Commissioning to quantify the impact of optical ghosts on the LSST data. 
In particular, we find that ${\sim}$ 0.57\% of the LSSTCam focal plane is impacted by optical ghosts when averaged across all bands.
We also use data from the Collimated Beam Projector to measure the reflectances of various optical elements, generally confirming estimates of $\sim$2\% from the systems engineering throughput predictions.  
\end{abstract}


\section{Introduction}
\begin{figure}[t!]
    \centering
    \includegraphics[width=0.6\textwidth]{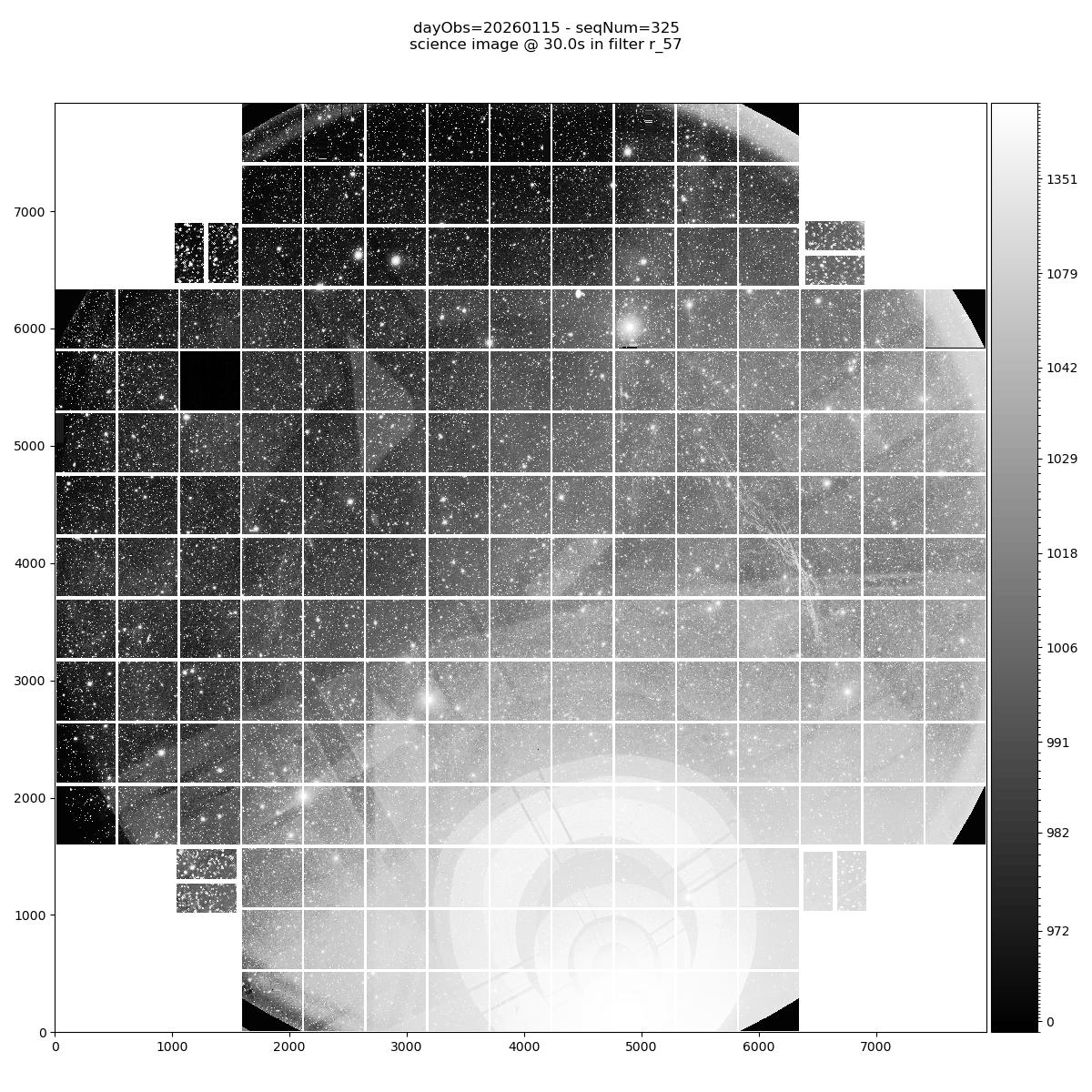}
    \caption{Post-ISR image of the LSSTCam focal plane (visit 2026011500325; {\it r} band) with optical ghosts created by a bright star (Canopus/Alpha Carinae) in the top right corner. } 
    \label{fig:optical_ghosts}
\end{figure}

Optical ghosts are caused by multiple reflections of light off of optical elements with non-zero reflectivity. 
Although ghosts are created by all astronomical sources, most of them are undetectable and have a negligible impact on astronomical measurements.
However, ghosts produced by bright stars can be impactful even if the reflectances of the optical elements are small ($\sim$1\%) because the total flux of photons from these stars are large. Figure~\ref{fig:optical_ghosts} shows an example of optical ghosts created by a bright star (Canopus/ Alpha Carinae) in the top right corner of the LSST Camera focal plane during commissioning of the Vera C.\ Rubin Observatory. %2025110500406. 

Ghosts produced by bright stars often pose challenges to sky background estimation and photometric measurements. 
They are also sources of contamination for low-surface-brightness science \citep[e.g.,][]{Tanoglidis:2021,Tanoglidis:2022}. 
Measuring the area of the focal plane affected by these artifacts is crucial to assess the impact on these science cases.

The LSST System Requirement\footnote{OSS Requirements: \url{https://docushare.lsst.org/docushare/dsweb/Get/LSE-30};  LSR Requirements: \url{https://docushare.lsst.org/docushare/dsweb/Get/LSE-29} } on the impact of optical ghosts states that the: ``Percentage of image area that can have ghosts with surface brightness gradient amplitude of more than 1/3 of the sky noise over 1 arcsec shall be less than 1\%''. 
The wording of this requirement leaves some room for interpretation. We opt to define the requirement somewhat conservatively as the total area of the focal plane containing a ghost with a flux that is greater than 1/3 of the sky noise on any detector. 
In order to satisfy this requirement, the total affected area should not exceed 1\%. 
We also assume that this requirement is intended to apply to the ensemble of visits rather than each individual visit.

To illustrate the difference between the stated requirement and our interpretation, we perform an example calculation using {\it r} band sky background and zeropoint values from \cite{SMTN-002}.
The gradient-based requirement stipulates that the ghost flux gradient must not exceed $\Delta = \sigma_{\rm sky}/3 \approx 10$\,ct\,pixel$^{-1}$ over 5 pixels (the pixel scale is $0.2$ arcsec per pixel), a condition that is typically only violated at the edges of pupil ghosts.
Our interpretation instead counts all pixels where the ghost flux exceeds $\sigma_{\rm sky}/3$, which is a more conservative criterion than the gradient-based requirement.
Even so, at this threshold the impact on the overall noise remains negligible since the sky flux ($\sim877$\,cts in {\it r} band) far exceeds the ghost flux.
The ghosts contribute a larger relative fraction of flux in {\it u} band, where the sky background is significantly lower ($\sim30$\, cts), making the ghost contribution more significant relative to the sky noise.

Measuring the fluxes of the ghosts with on-sky data to estimate the ghost-flux-to-sky-noise ratio is challenging. 
Uneven flat-fielding during commissioning led to gradients in the mean sky counts across the focal plane, affecting the sky noise and ghost flux across the ghost. 
Most of the prominent ghosts were also found to be spatially degenerate, with some ghosts fully overlapping others. 
This made fitting the amplitude of each individual ghost difficult.
To circumvent these challenges, we chose instead to calculate the expected impact of ghosts by performing ray-tracing simulations of known bright stars, taking into account their positions relative to the boresight and approximate brightnesses, as well as the locations of Rubin commissioning observations. 
The simulated fluxes of the ghosts were compared and calibrated to the Rubin commissioning data to validate this approach.

We use the Collimated Beam Projector (CBP) -- a 30\,cm Schmidt telescope refitted with a laser -- to produce ghosts on the focal plane.
We measure the fluxes of these ghosts and derive estimates for the reflectances of the optical elements (Filter, Detector \& Lenses) of LSSTCam.
These measurements serve as validation of the reflectances used in the simulations where we estimate the area of sky impacted by optical ghosts.

This document is structured as follows. 
Section~\ref{sec:ghost_impact} delineates the procedure and results of our estimation of the sky area impacted by ghosts.
Section~\ref{sec:reflectances} shows our measurements of the optical reflectances of the camera optics using ghosts.

\section{Determining Ghost Impact}
\label{sec:ghost_impact}
This section describes the steps used to estimate the ghost-impacted area for an individual LSSTCam visit. 
This process was repeated for each exposure in the Rubin commissioning to derive a quantitative assessment of the ghost-impacted area in the science verification surveys.

\begin{figure}[h]
    \centering
\includegraphics[width=0.45\textwidth]{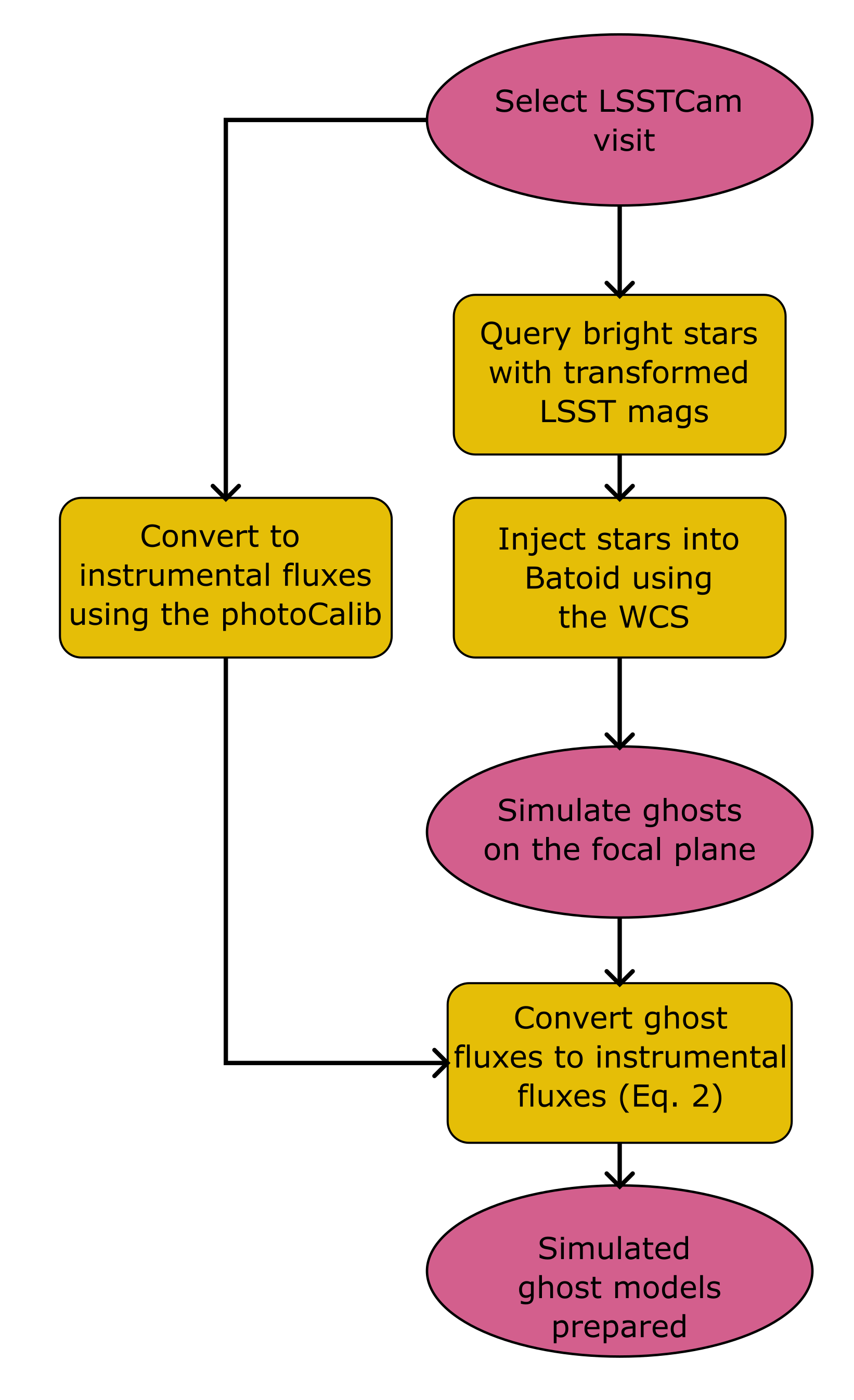}
    \caption{Flowchart showing the procedure used to generate optical ghost templates and surface brightness estimates.} 
    \label{fig:ghost_model_flowchart}
\end{figure}

\subsection{Bright Stars}
For each Rubin exposure, we retrieve all bright stars from the Yale Bright Star Catalog \citep{Hoffleit1991} within 1.9 deg of the Rubin boresight.
We transform the $V$-band magnitudes of these bright stars in the LSST bands by first transforming into the DES photometric system ($grizY$) following the procedure described in Appendix B of \cite{2021ApJS..255...20A}.
We then follow the approach of \cite{DMTN-277} to transform from the DES system to predicted LSST magnitudes. 

In order to accomplish the first step in this task, we use the transformation equations listed in Eq.~\ref{eq:ybsc_to_des} to transform the magnitudes of the stars using the $V$ magnitude and the $B-V$ colour to DES magnitudes. These transformations are valid for -0.2 $<B-V\leq $ 2.2.
For stars that do not have $B-V$ colour or stars whose colours lie outside of the range above (113 out of 8673 stars), we use the $V$ band magnitude in place of transformed LSST magnitudes for the simulations.

\begin{equation}
\label{eq:ybsc_to_des}
    \begin{aligned}
        g_{\small \rm DES} = V + 0.496 (B-V) -0.07  \\
        r_{\small \rm DES} = V  -0.543 (B-V) + 0.128  \\
        i_{\small \rm DES} = V -1.04 (B-V) +0.312 \\
        z_{\small \rm DES} = V -1.302 (B-V) +0.417 \\
        Y_{\small \rm DES} = V -1.416 (B-V) +0.504 
    \end{aligned}
\end{equation}

To transform from DES to LSST, we use transformation equations derived in the construction of The Monster \citep{DMTN-277}. 
We are unable to use this sequence of transformations to generate LSST {\it u} band magnitudes because we did not have transformation equations from the DECam {\it u} to LSSTCam {\it u}.  
In this case, we use the $V$ magnitudes instead of the transformed LSST {\it u} magnitudes. 
We use the $V$ band magnitude for all star magnitudes that we were unable to transform to LSST magnitudes.

\subsection{Simulations}
\label{sec:sims}
In this section we describe the use of the \texttt{Batoid} Python package \citep{Batoid} to generate simulated optical ghosts from bright stars in a particular visit. 

\subsubsection{Pipeline}
\label{sec:pipeline}
\begin{figure}[h]
    \centering
    \includegraphics[width=0.5\textwidth]{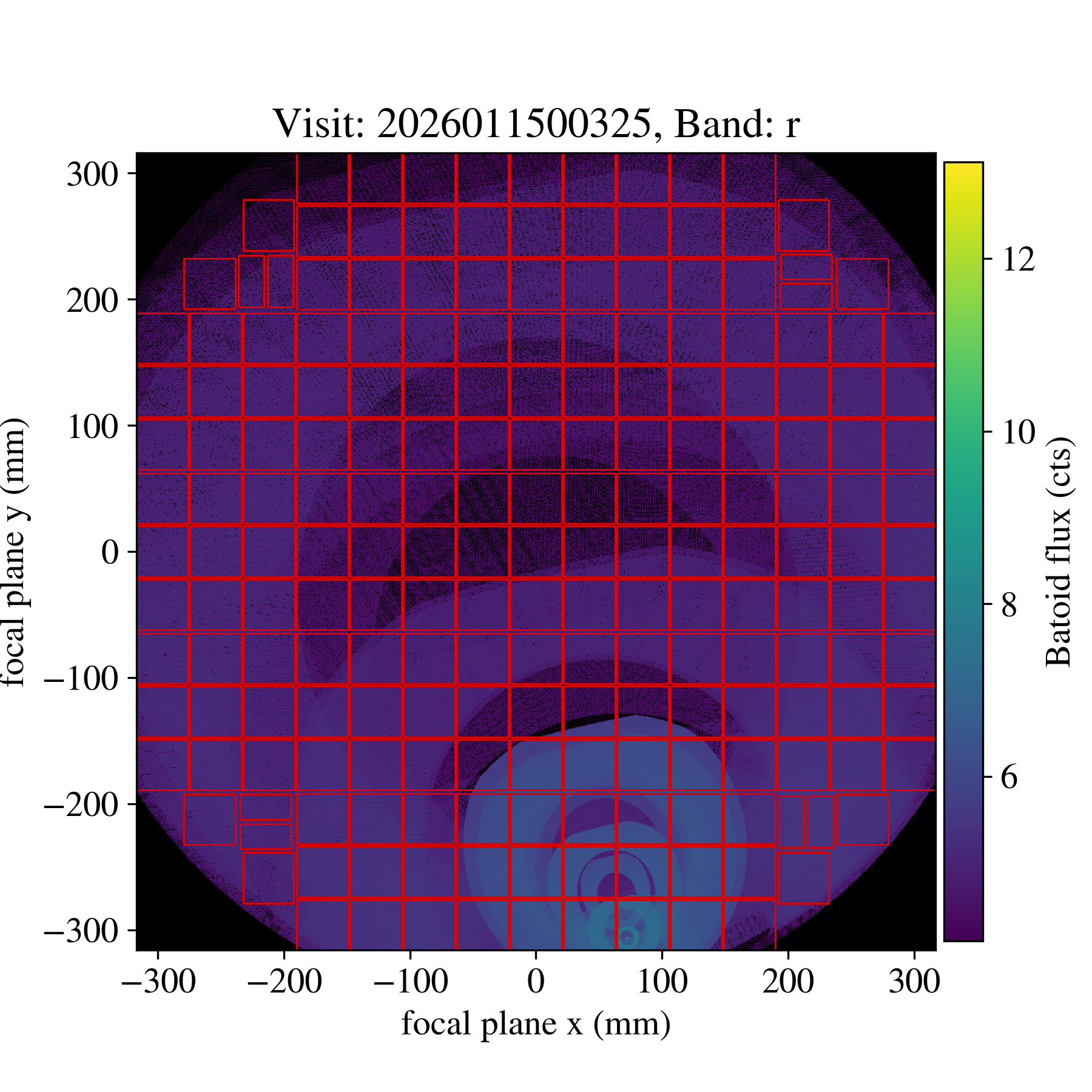}
    \caption{Stacked simulated ghosts produced by a \texttt{Batoid} simulation using the procedure delineated in Section~\ref{sec:pipeline}.} 
    \label{fig:stacked_ghosts}
\end{figure}

Figure~\ref{fig:ghost_model_flowchart} shows a flowchart of the pipeline used to generate morphology, flux, and surface brightnesses of ghosts. 
The optical reflectances of the filters and lenses in the simulation were set to the values produced by the systems engineering simulations\footnote{\url{https://github.com/lsst-pst/syseng_throughputs/blob/main/notebooks/Components.ipynb}}. 
The bright stars were initialized into the simulation using the 
\\ \texttt{preliminary\_visit\_image.wcs} object. 
The rays were then propagated through the full optical model. 
The flux of each ghost was converted from arbitrary flux units in \texttt{Batoid} to instrumental flux using Eq.~\ref{eq:batoid_flux_to_instflux}, where $f_i$ is the instrumental flux of a particular ghost, $\phi_{\rm tot}$ is the total flux of all ghosts in arbitrary flux units, $\phi_{\rm src}$ is the flux of the star in arbitrary flux units and $f_{\rm src}$ is the instrumental flux of the star. 
$f_{\rm src}$ is calculated by converting the magnitude of the star in the LSST band to instrumental fluxes using the \texttt{preliminary\_visit\_image.photoCalib} object. 
\begin{equation}
\label{eq:batoid_flux_to_instflux}
    f_{i} = \frac{\phi_i}{\phi_{\rm tot}+\phi_{\rm src}} f_{\rm src}
\end{equation}
Figure~\ref{fig:stacked_ghosts} shows the stacked optical ghosts produced by the simulation for visit 2026011500325.

\subsubsection{Ghost Nomenclature \& Morphology}
\label{sec:nomenclature}
In this document, ghosts are labeled by the two optical elements that created them, ordered by the sequence of reflections. 
`L\#' stands for lens (\# corresponds to the lens number which can be 1, 2 or 3), `F' for filter, and `D' for detector. 
The 1 or 2 at the end of each alphanumeric sequence denotes the surface of the optical element that the ray bounced off of. 
For example, the `L31-F2' ghost was created due to the reflection of rays from the first surface (entrance) of L3 and again from the second surface (exit) of the Filter.

\begin{figure}[h!]
    \centering

\includegraphics[width=1\textwidth]{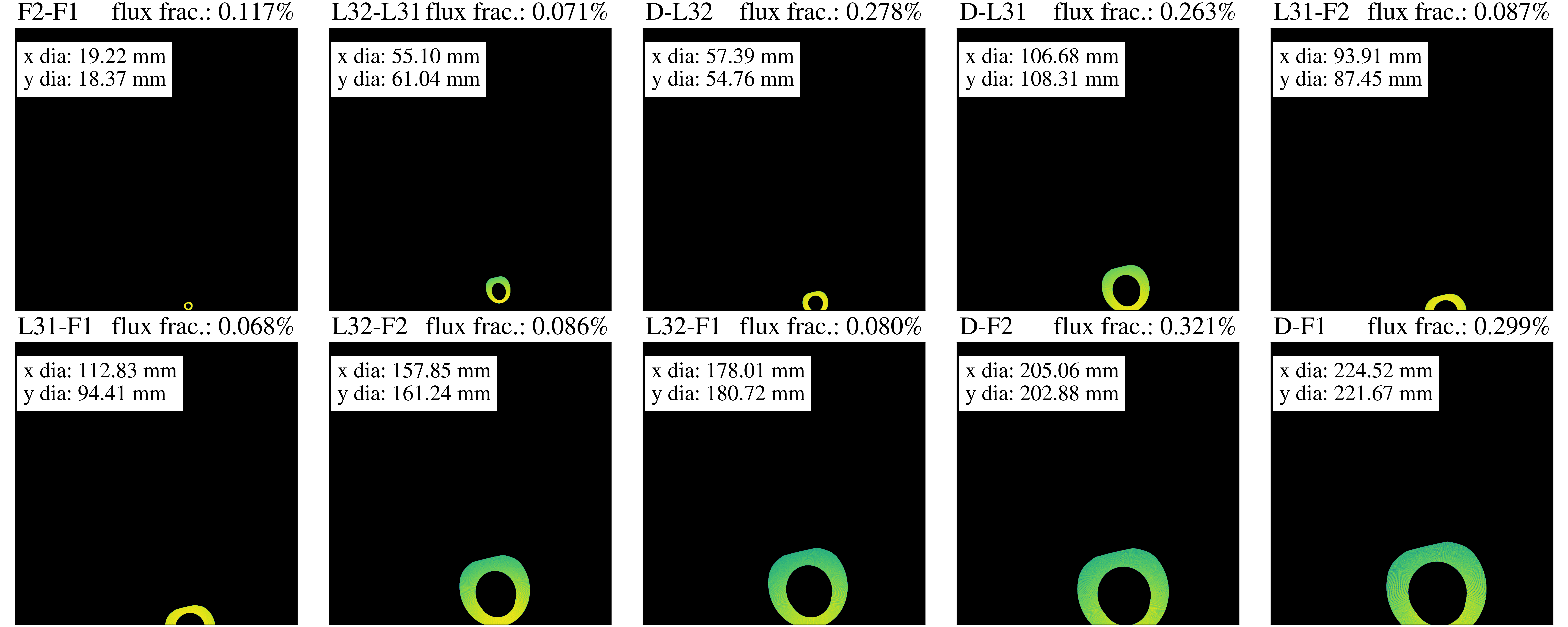}
    \caption{The ten most commonly occurring ghosts mentioned in Section~\ref{sec:nomenclature} from the stacked simulation in Fig.~\ref{fig:optical_ghosts}. 
    Each ghost is labelled by the pair of optical elements that the ray reflected off of, the fraction of stellar flux that contributed to the ghost and the x and y diameter.} 
    \label{fig:optical_ghosts_individual}
\end{figure}

The ten most commonly occurring ghosts are: F2-F1, L32-F1, L32-F2, L31-F1, L31-F2, L32-L31, D-F1, D-F2, D-L31, D-L32. 
The F2-F1 ghost is the smallest and most stable with respect to the position on the focal plane, while the D-F1 \& D-F2 ghosts are the biggest and have the largest changes in their footprint aspect ratio as the star moves off-axis.

Figure~\ref{fig:optical_ghosts_individual} shows an example LSSTCam visit with optical ghosts from Canopus/ Alpha Carinae, along with a few of the simulated ghosts labeled by the optical elements responsible for creating them, their size, and the fraction of stellar flux that contributes to them.

\subsection{Measuring the Impacted Area}

\begin{figure}[h!]
    \centering

\includegraphics[width=0.45\textwidth]{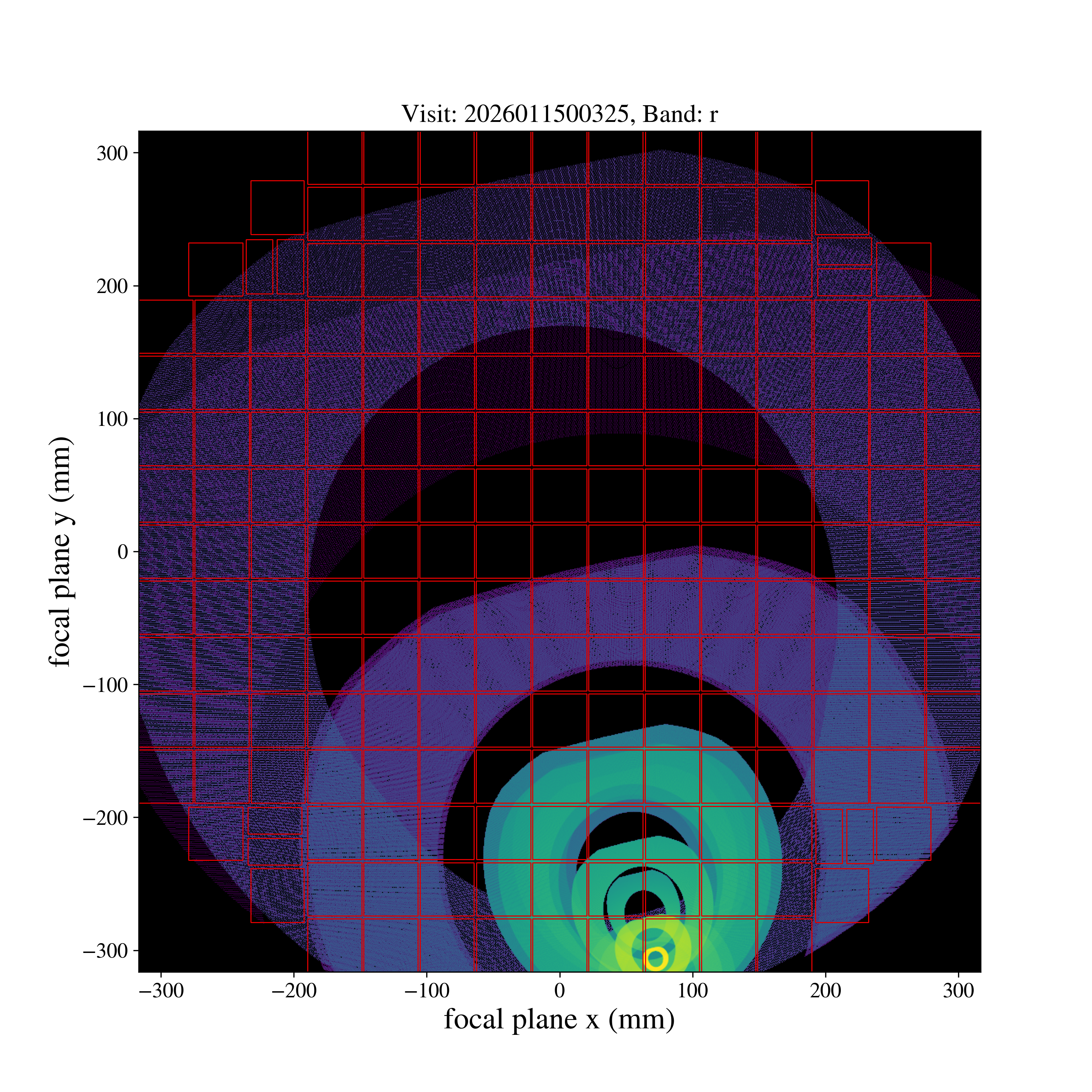}
    \caption{Area of the focal plane marked as impacted by optical ghosts in green for LSSTCam visit 2026011500325. 
    For this particular visit, we find the area significantly impacted by ghosts to be 56.7\% of the total focal plane area.} 
    \label{fig:impacting_ghosts}
\end{figure}

To measure the impacted area as defined by our interpretation of the system requirement, we first generate individual ghost models using the algorithm in Section~\ref{sec:sims}. 
For each individual ghost, the ghost area (in pixels) and the total ghost flux (in counts) is calculated per detector. 
The surface brightness of the ghost ($S_{\rm n}$) is then calculated by dividing the total ghost flux per detector ($f_{\rm tot,n}$) by the ghost area per detector ($A_{\rm n}$), in units of counts per pixel as shown in Eq.~\ref{eq:surface_brightness}. 
Here, {\it n} labels the detector number.  
\begin{equation}
\label{eq:surface_brightness}
    S_{\rm n} = \frac{f_{\rm tot,n}}{A_{\rm n}}
\end{equation}

The median measured sky noise per detector is queried for the visit through the LSSTCam ConsDB \citep{DMTN-227}. 
We calculate the ratio of the ghost surface brightness to the sky noise. 
If this ratio exceeds 1/3, the ghost area is marked as ``impacted''. 
Repeating this process for each ghost and accounting for ghost overlap gives us the final area of the focal plane that is impacted by ghosts.
Note that in this procedure, we apply the ratio threshold of 1/3 to each ghost individually, i.e.\ the flux contributions to an individual detector from multiple ghosts are \textit{not} integrated prior to thresholding.
Figure~\ref{fig:impacting_ghosts} shows the area impacted by ghosts measured for visit 2026011500325. 
The impacted area is 7.0 sq. deg., which is 56.7\% of the total focal plane area.

\subsubsection{Impacted Area Statistics}
\label{sec:stats}
\begin{figure}[h!]
    \centering
\includegraphics[width=0.6\textwidth]{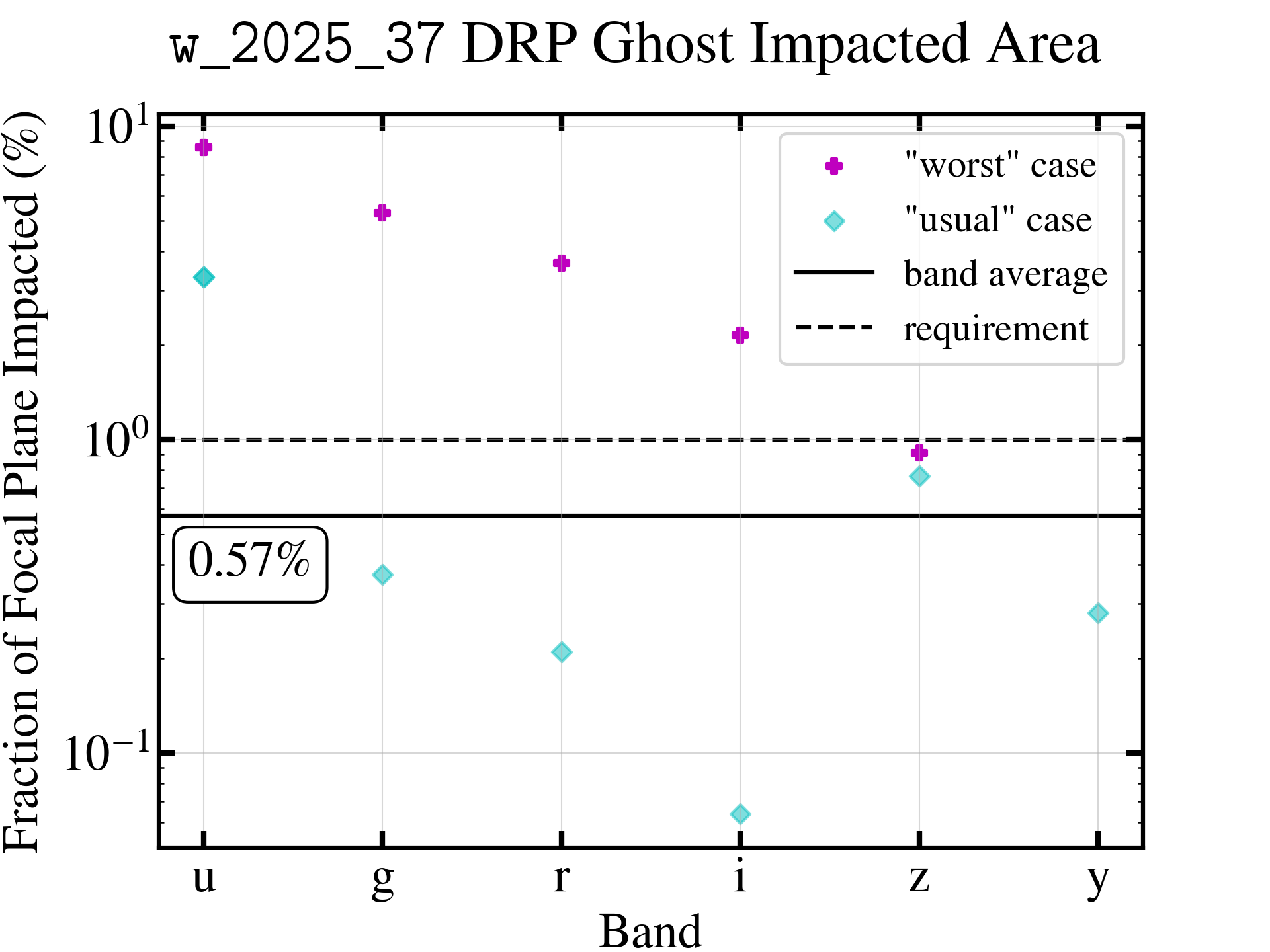}
    \caption{Area of the focal plane impacted by ghosts separated by band in the w37 DRP (cyan) and in a set of highly-ghosted visits in all bands except {\it y} (magenta). 
    The impacted area in the DRP is 0.57\% when averaged across all bands. The highest contribution is from {\it u} (which is driven primarily by the low sky background) and the lowest contribution is from {\it i}.} 
    \label{fig:drp_ghost_stats}
\end{figure}

We run the pipeline on two datasets to produce an estimate for the ``usual" case and ``worst" case ghost-impacted area. 
We used the weekly 37 intermittent DRP as the ``usual'' scenario to represent the LSST wide-fast-deep survey. The DRP covers a $30^\circ \times 20^\circ$ region with a centroid RA, Dec.\ $\sim(311.45^\circ, -18.43^\circ)$. 
We ran the pipeline above to measure the ghost-impacted area of the focal plane on $\sim$3100 visits in the DRP.
For the ``worst'' case estimate, we ran our pipeline on 20 visits in each band (except {\it y}, for which 20 visits with severe ghosting could not be identified) that were selected visually to contain large amounts of ghosting. 
Figure~\ref{fig:drp_ghost_stats} shows that the ghost-impacted area in the DRP averaged over all bands and visits is 0.57\%. 
When separated by band, the average impacted area in {\it u} is the highest at $\sim$8\% and smallest in {\it r} ($\sim$0.2\%) and {\it i} ($<$0.1\%). 
The anomalously high value in {\it u} is likely due to the lower sky background and a much higher filter reflectance in the {\it u} band.
Table~\ref{tab:impacted_area} shows the total impacted area and surveyed area in each band in the w37 intermittent DRP. 
\begin{table}[h!]
    \centering
    \begin{tabular}{lcc}
        \hline
        Band & Impacted Area [deg$^{2}$] & Surveyed Area [deg$^{2}$] \\
        \hline
        {\it u} & 68.95 & 2225.91 \\
        {\it g} & 14.01 & 4897.00 \\
        {\it r} & 9.94 & 6084.15 \\
        {\it i} & 4.04 & 9719.80 \\
        {\it z} & 33.56 & 8668.68 \\
        {\it y} & 7.96 & 6603.53 \\
        \hline
    \end{tabular}
    \caption{Intermittent DRP (w37) ghost-impacted area and total surveyed sky area separated by band. \label{tab:impacted_area}}
\end{table}

\begin{figure}[h!]
    \centering
\includegraphics[width=0.9\linewidth]{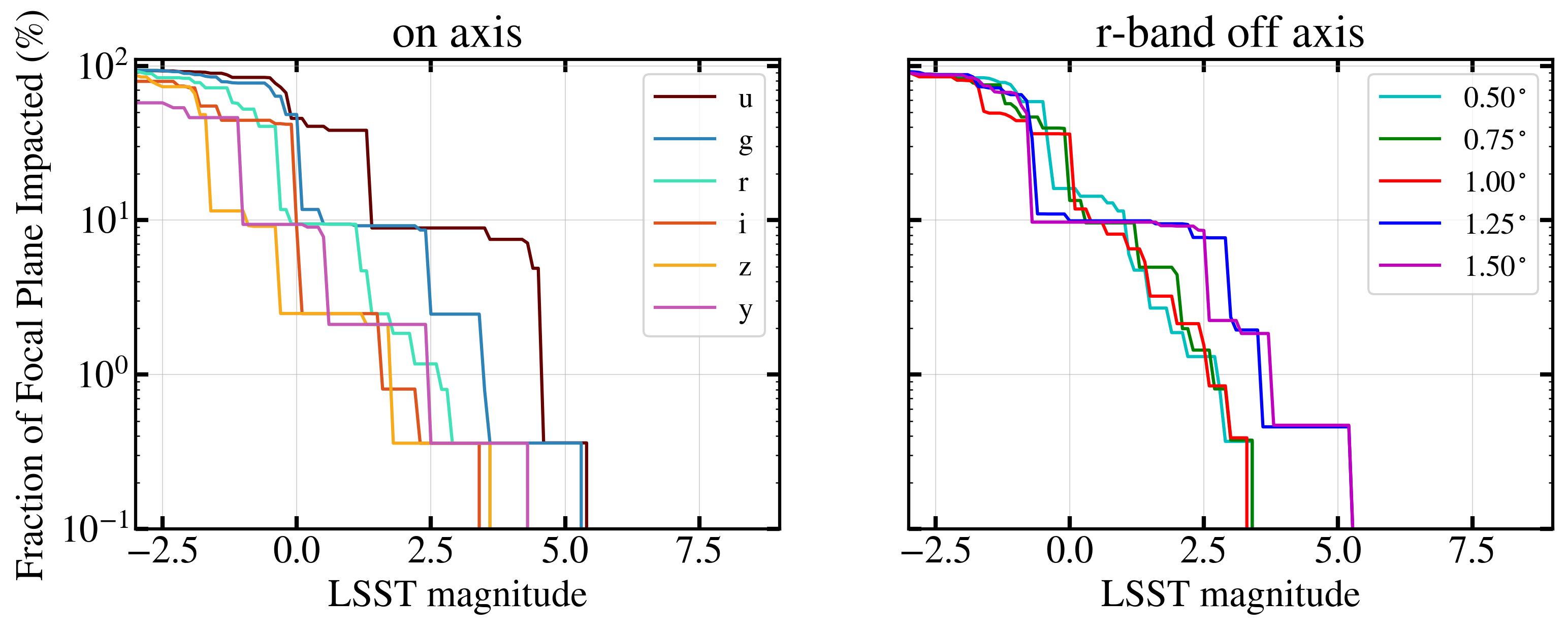}
    \caption{(Left) Fraction of the focal plane impacted by optical ghosts as a function of star magnitude for a single on-axis simulated star. 
    (Right) The r-band impacted area as a function of star magnitude for different off-axis positions of the simulated star.} 
    \label{fig:drp_ghost_impact_sim}
\end{figure}

We also simulate a single star with different magnitudes at various off-axis positions on the focal plane to plot the impacted area as a function of star magnitude in Fig.~\ref{fig:drp_ghost_impact_sim}.
The left panel shows the impacted area as a function of the magnitude of an on-axis bright star separated by band, and the right panel shows the weak dependence of the impacted area on the star's offset angle relative to the boresight. 

\begin{figure}[h!]
    \centering
\includegraphics[width=1\textwidth]{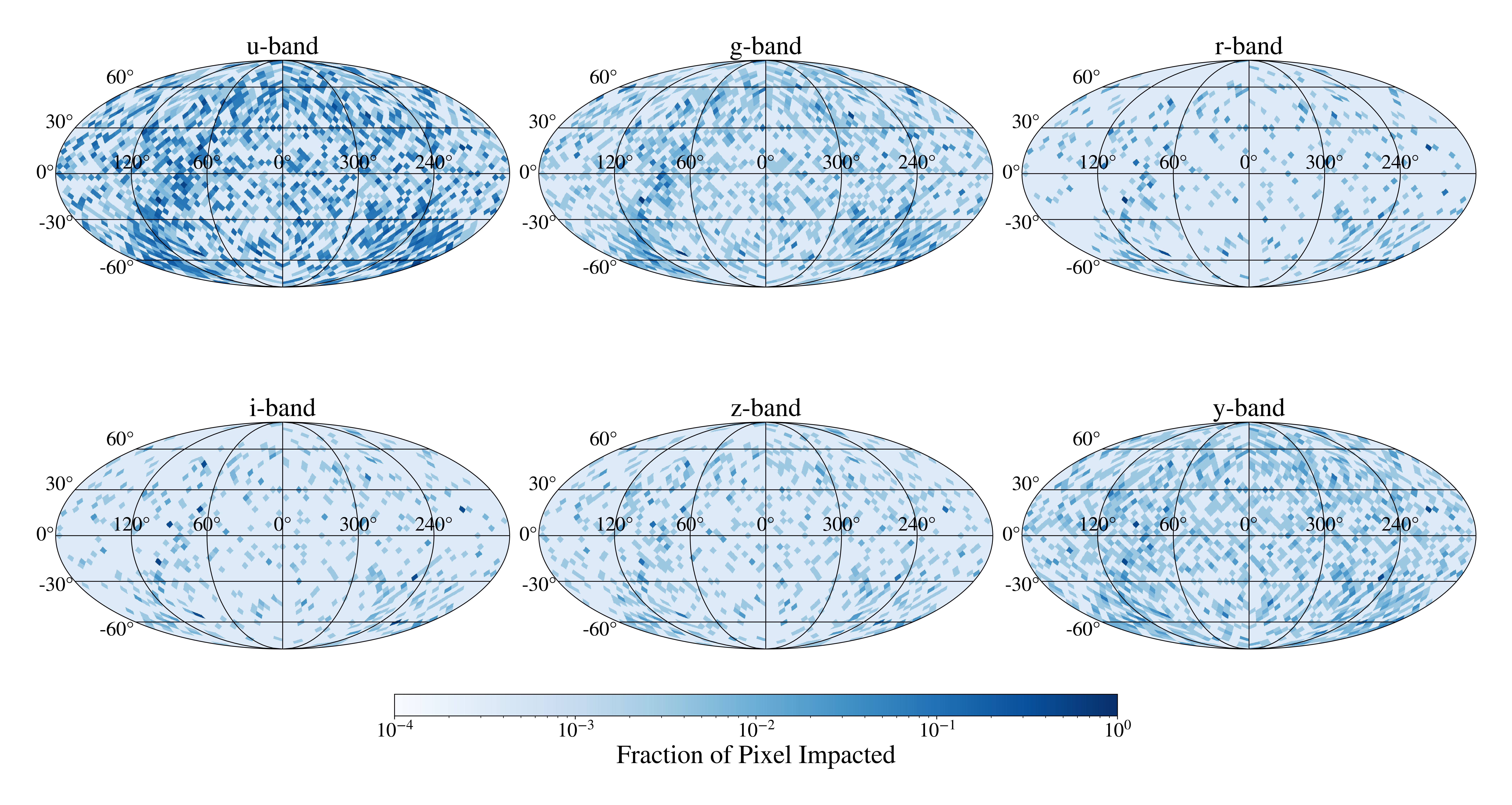}
    \caption{Sky maps showing the fraction of ghost-impacted area in each pixel from the stars in the Yale Bright Star Catalog, separated by band using the simulations in Fig.~\ref{fig:drp_ghost_impact_sim}.} 
    \label{fig:drp_ghost_sky_maps}
\end{figure}

We use the left panel from Fig.~\ref{fig:drp_ghost_impact_sim} as a look-up table to generate a sky map of the impacted area from all the stars in the Yale Bright Star catalog, assuming that each star was observed on-axis. 
We use the nominal zeropoints and dark sky counts from  \cite{SMTN-002} to calculate the sky noise in each band. We also assume that the exposure times for each band were 30\,s. 
These maps for each band are shown in Fig.~\ref{fig:drp_ghost_sky_maps}. 

Note that in these maps, the impacted area from each bright star within the same pixel is simply added. 
This leads to a slight over-counting of the impacted area, as stars that are close to each other can produce overlapping ghosts, changing the total impacted area of the two stars from a simple sum to a union of the individual impacted areas.

\section{Measuring Reflectances of Optical Elements}
\label{sec:reflectances}
In this section, we describe the procedure used to measure the reflectances of the LSSTCam optical elements (i.e., Detector, Filter, L1, L2, L3) using the strength of optical ghosts produced in test data collected with the Collimated Beam Projector (CBP) \citep{2016SPIE.9906E..0OI}. 
This serves as a verification of the reflectances used in the \texttt{Batoid} simulations in the previous sections, which in turn also validates the reflectances used in systems engineering simulations.

\subsection{Optical Ghosts and Reflectances}
\label{sec:reflectance_theory}
To form a first-order ghost, a ray of light must be reflected successively by two optical elements (higher-order ghosts also exist, but are highly suppressed, as they require four or more successive reflections).
The probability that a ray contributes to the flux of the ghost is $P_{AB} = R_A R_B$, where $R_A$ and $R_B$ are the reflectances of the optical elements A and B, respectively.
The flux of an optical ghost can therefore be related to the flux of the source through $P_{AB}$, as shown in Eq.~\ref{eq:ghost_flux_prob}, where $f_{A, B}$ is the flux of the ghost and $f_{\rm src}$ is the total incident flux of the astronomical source (i.e., bright star).

\begin{equation}
\label{eq:ghost_flux_prob}
f_{A,B} = R_A R_B f_{\rm src}
\end{equation}

\noindent Measurements of $f_{A,B}$ and $f_{\rm src}$ allow a determination of the combined reflectances of the optical elements involved in the production of the ghost.

Considering all ghosts produced by different combinations of optical elements leads to a set of coupled equations.
A subset of these equations, limited to ghosts arising from a reflection off of a filter surface (F1 or F2) and either the  Detectors (D) or a surface of Lens 3 (L31 or L32), is shown in Eq.~\ref{eq:coupled_eqs}.

\begin{equation}
\label{eq:coupled_eqs}
\begin{aligned}
f_{F1 , F2} &= R_{F1} R_{F2} f_{\rm src} \\
f_{L{32} , F1} &= R_{L{32}} R_{F1} f_{\rm src} \\
f_{L{32} , F2} &= R_{L{32}} R_{F2} f_{\rm src} \\
f_{L{31} , F1} &= R_{L{31}} R_{F1} f_{\rm src} \\
f_{L{31} , F2} &= R_{L{31}} R_{F2} f_{\rm src} \\
f_{D , F1} &= R_{D} R_{F1} f_{\rm src} \\
f_{D , F2} &= R_{D} R_{F2} f_{\rm src}.
\end{aligned}
\end{equation}
\noindent This system of equations can be linearized in log-space as follows:
\begin{equation}
    \label{eq:log-coupled-eqs}
    \log\left(\frac{f_{A,B}}{f_{\rm src}}\right) = \log R_A + \log R_B.
\end{equation}
With a redefinition of variables $x_A \equiv \log R_A$ and $b_{A,B} \equiv \log \left(\frac{f_{A,B}}{f_{\rm src}}\right)$, the system of equations in log-space can be written as a matrix problem $A\vec{x} = \vec{b}$ (expanded form in Eq.~\ref{eq:reflectance_matrix}), where the reflectances can be solved for with least squares by inverting the matrix $A$. 
Here, the vector $\vec{x}$ contains the log reflectances and the vector $\vec{b}$ contains the ghost fluxes.
This system can thus be solved independently of knowing the source flux.
\begin{equation}
\label{eq:reflectance_matrix}
\begin{pmatrix}
1 & 1 & 0 & 0 & 0 & \dots \\
0 & 0 & 0 & 0 & 0 & \\
1 & 0 & 0 & 0 & 0 &  \\
0 & 1 & 0 & 0 & 0 & \dots \\
\vdots & & & & \vdots & \ddots \\
\end{pmatrix} \, 
\begin{pmatrix}
x_{F_1} \\
x_{F_2} \\
x_{L_{11}} \\
x_{L_{12}} \\
\vdots \\
\end{pmatrix}\,\,
=
\begin{pmatrix}
b_{F_1 , F_2} \\
b_{L_{32} , L_{31}} \\
b_{L_{32} , F_1} \\
b_{L_{32} , F_2} \\
\vdots \\
\end{pmatrix}
\end{equation}

\subsection{Measuring Ghost Fluxes}
\subsubsection{Data}

\begin{figure}[h!]
    \centering
\includegraphics[width=0.2\textwidth]{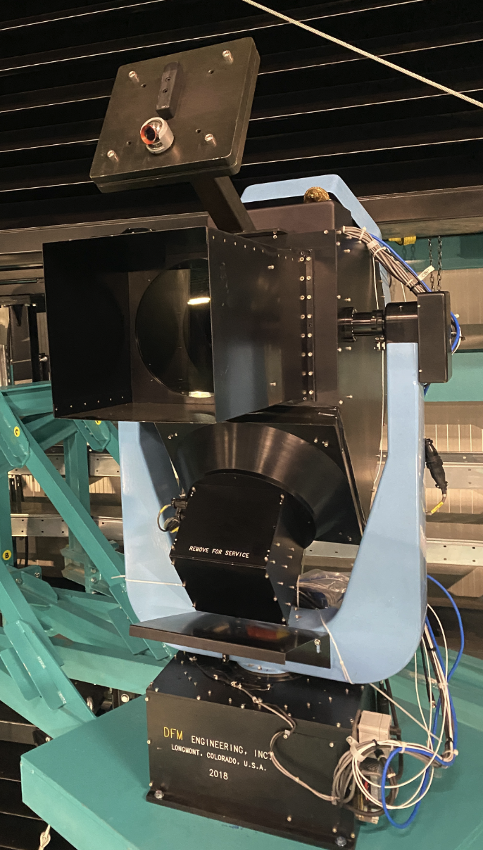}
    \caption{An image of the Collimated Beam Projector (CBP) from \cite{SITCOMTN-152}.} 
    \label{fig:cbp}
\end{figure}

The Collimated Beam Projector (CBP), shown in Fig.~\ref{fig:cbp}, is a repurposed 30\,cm Schmidt telescope that has been fitted with a laser and mounted to the Rubin dome. It is located approximately 12.5\,m from the Telescope Mount Assembly (TMA) at a height of approximately 14\,m. 
The CBP projects a monochromatic point-like source in a parallel beam onto the primary mirror (M1), thereby producing an artificial light source mimicking the behaviour of a star.
We use a mask with a single pinhole of size \SI{1}{\milli\m}, which is then projected onto the focal plane with $\sim16\times$ magnification.
Note that the CBP does not illuminate the entirety of M1 as a star would, which changes the morphology of the observed ghosts.

\begin{figure}[h!]
  \centering

  \begin{subfigure}[t]{0.45\textwidth}
    \centering
    \includegraphics[width=\linewidth]{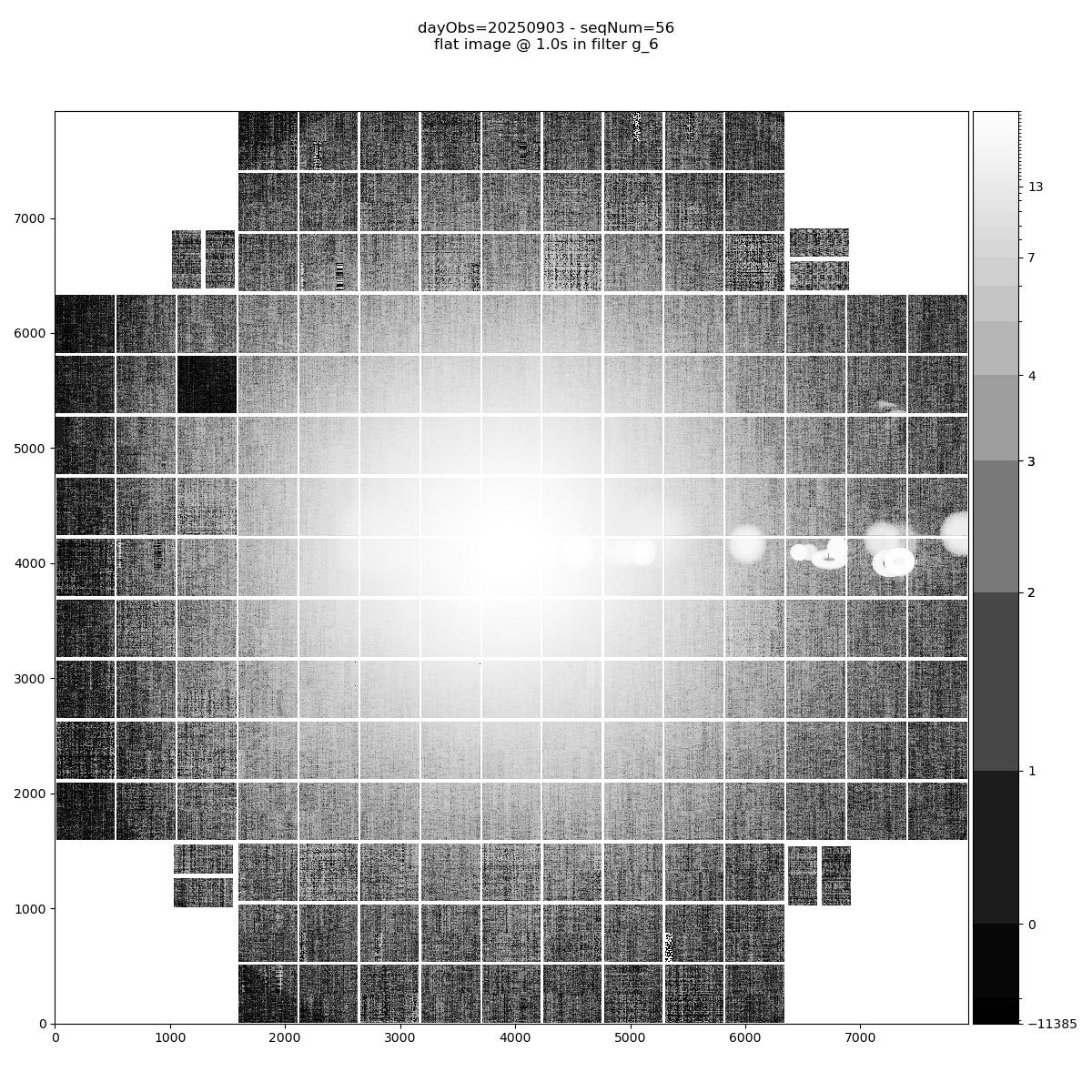}
    \caption{}
    \label{subfig:cbp_ghosts_a}
  \end{subfigure}
  \hfill
  \begin{subfigure}[t]{0.45\textwidth}
    \centering
    \includegraphics[width=\linewidth]{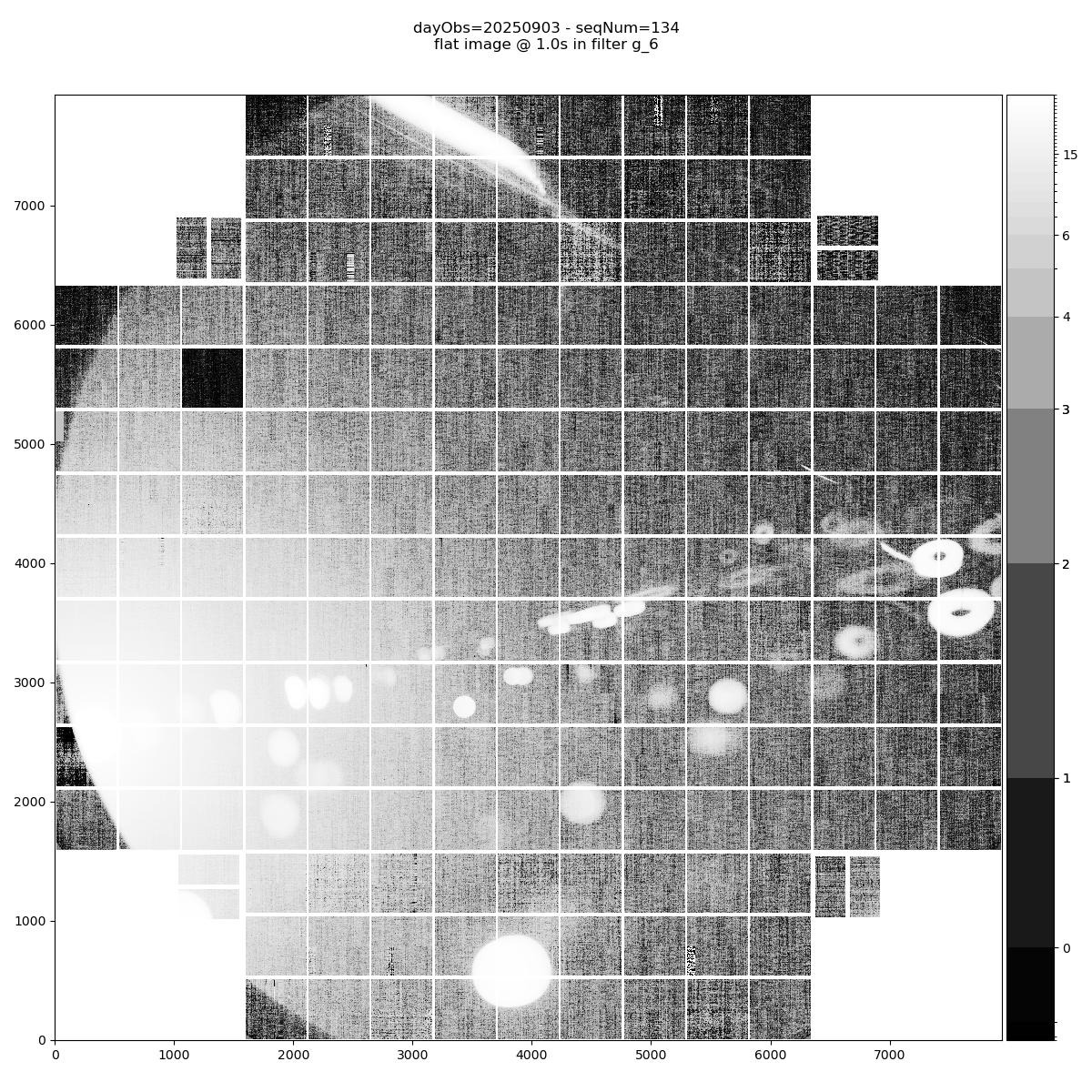}
    \caption{}
    \label{subfig:cbp_ghosts_b}
  \end{subfigure}

  \caption{Subfigures~\ref{subfig:cbp_ghosts_a} and \ref{subfig:cbp_ghosts_b} show the ghosts produced by a single spot CBP beam at two different pointings of the CBP in the {\it g} band.}
  \label{fig:cbp_ghosts}
\end{figure}

Figure~\ref{fig:cbp_ghosts} shows two examples of the CBP ghost data collected at various CBP-TMA pointings in the \textit{g} band.
We use the CBP to generate optical ghosts on the focal plane to avoid the challenges associated with on-sky data, including sky background fitting and crowded stellar fields.
Furthermore, the flexibility in the CBP--TMA pointing configuration allows us to generate ghosts that are more spatially separated than those that come from on-axis, on-sky data (cf. Figs.~\ref{fig:optical_ghosts} and \ref{fig:cbp_ghosts}).

Our data was collected during one CBP run performed during Rubin Observatory commissioning on 2025-09-03.
Our data taking configuration involved sweeping the CBP leftward (toward $-X$ in focal plane coordinates) starting from the center of the focal plane. 
We collected 36 images with `no filter' (519\,nm), 36 images in \textit{g} band (520\,nm), 18 in \textit{r} band (620\,nm) and 7 in \textit{i} band (760\,nm) (visit numbers 2025090300038 to 2025090300133).
The chosen wavelengths lie around the middle of their respective bandpasses.

\subsubsection{Ghost Flux Fitting}
\label{sec:ghost_fit}

\begin{figure}[h!]
  \centering

  \begin{subfigure}[t]{0.54\textwidth}
    \centering
    \includegraphics[width=\linewidth]{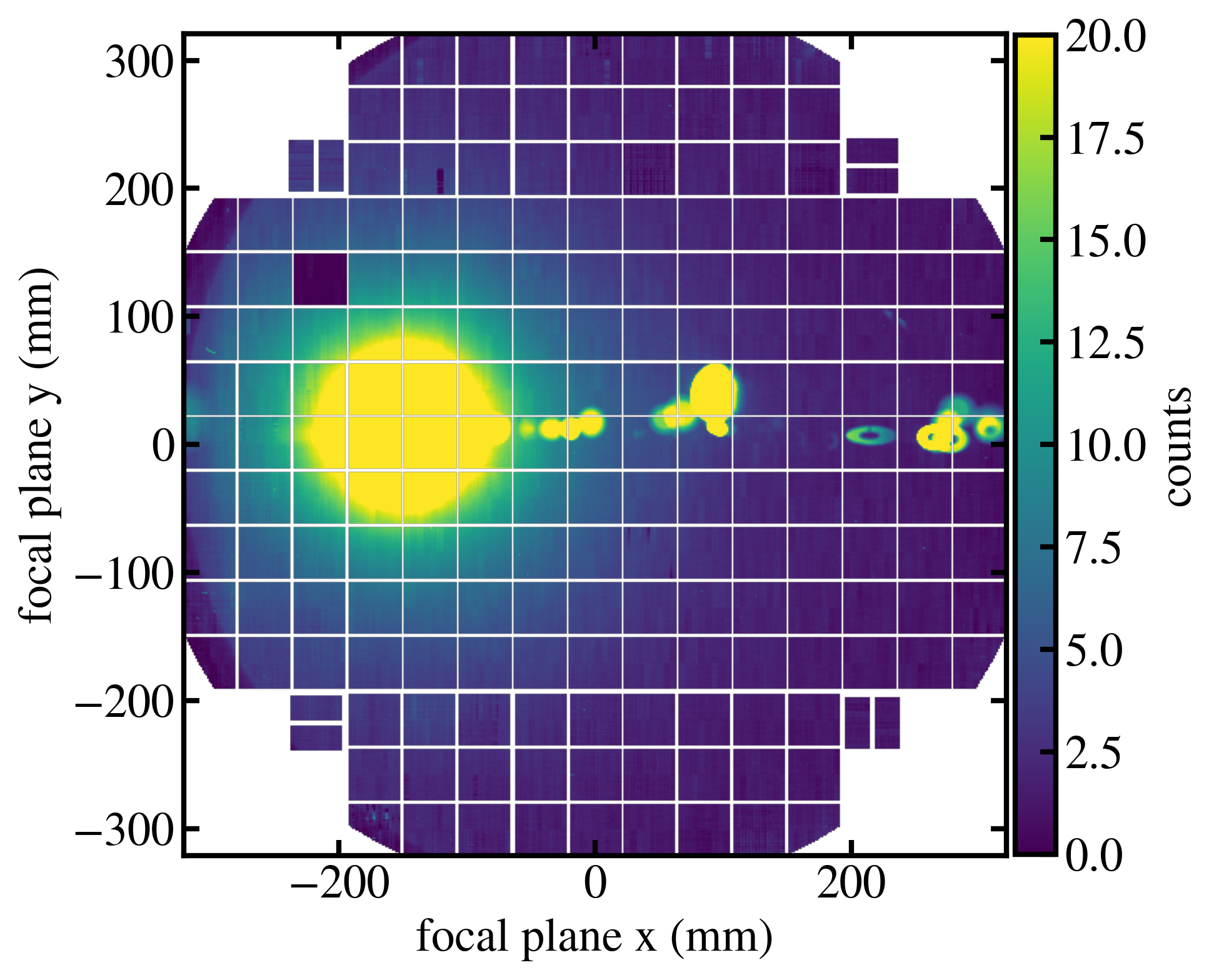}
    \caption{}
    \label{subfig:cbp_data}
  \end{subfigure}
  \hfill
  \begin{subfigure}[t]{0.45\textwidth}
    \centering
    \includegraphics[width=\linewidth]{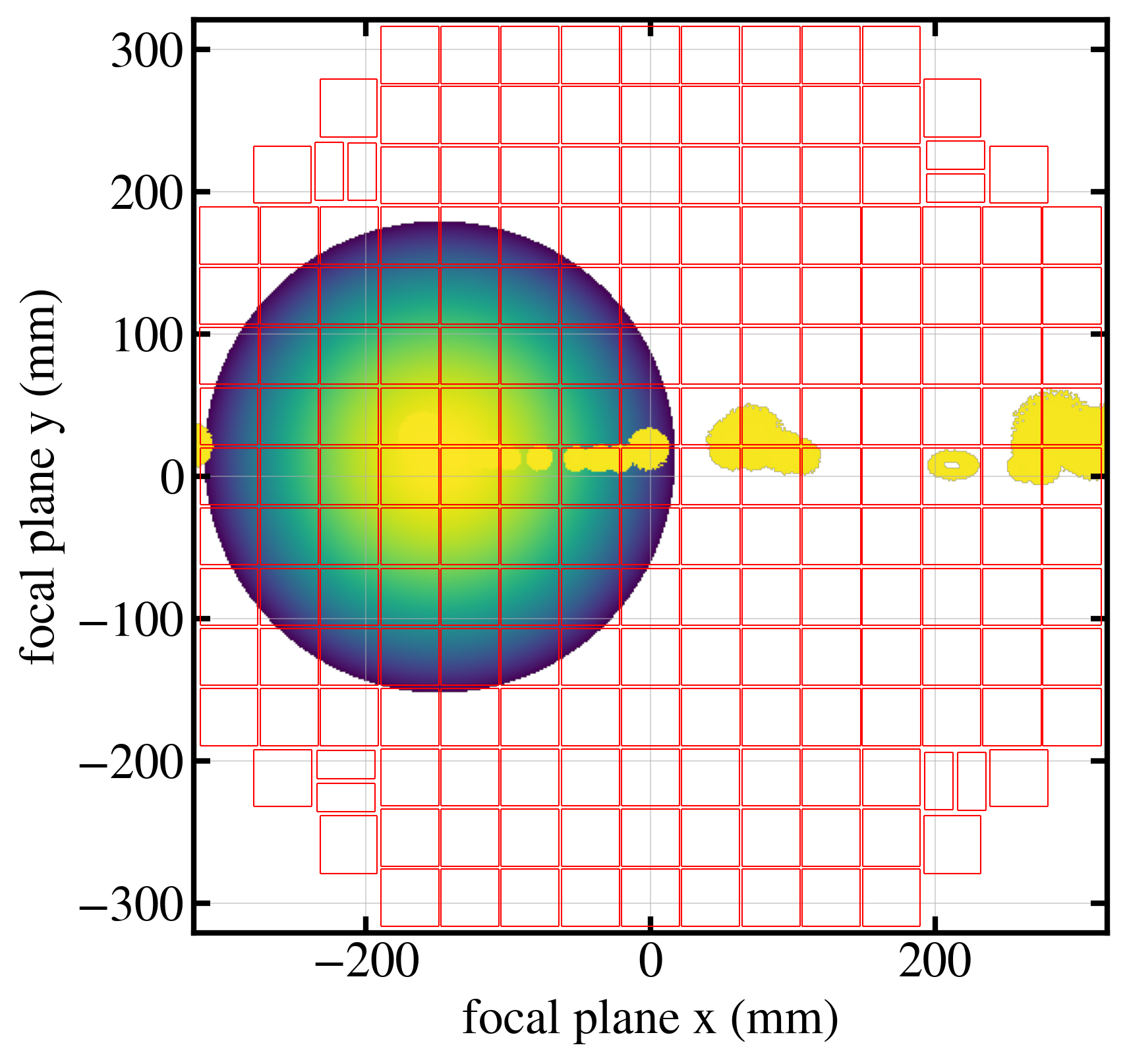}
    \caption{}
    \label{subfig:cbp_sim}
  \end{subfigure}

  \caption{The left panel shows an example of a CBP ghost image (visit 2025090300060) in {\it g} band. 
  Note that the colour bar is clipped to show the faint ghosts; the source beam is on the order of $10^5$ counts. 
  The right panel shows the CBP ghosts simulated in \texttt{Batoid} along with the Gaussian template, all normalized to 1.}
  
  \label{fig:cbp_ghost_templates}
\end{figure}

We bin the data collected from the CBP into images of 494 $\times$ 493 pixels.
We use \texttt{Batoid} to generate templates of the CBP ghosts using its built-in CBP model. 
Each simulated ghost is binned with the same resolution as the data, and then the template is normalized to have a maximum amplitude of 1.
We also include a constant background template.
Finally, we use a flat-top Gaussian model as described in Eq.~\ref{eq:flat_gaussian} to account for the saturated source beam.
Here, $r_0$ is the radius within which the Gaussian is "saturated" at amplitude $A$.
We use a radial average of the data (centered at the beam location on the focal plane retrieved from \texttt{Batoid}) to determine the width and amplitude of the Gaussian model.
The background and Gaussian templates are also normalized to a maximum value of 1.
We find that the following 16 ghosts originating from the detectors (D), filter (F), and lenses (L1, L2, L3) are the most prominent: L12-L11, L22-L11, L22-L12, F2-L21, F1-L21, F1-L22, L32-L22, L31-L22, D-L22, D-F1, D-F2, D-L31, L31-F2, L32-F1, L32-F2, L31-F1.
Figure~\ref{fig:cbp_ghost_templates} shows an example of the binned data and the 16 ghost templates and the Gaussian template simulated in \texttt{Batoid}.

\begin{equation}
\label{eq:flat_gaussian}
f(r) =
\begin{cases}
A, & r \le r_0 \\
A \exp\left(-\dfrac{(r - r_0)^2}{2\sigma^2}\right), & r > r_0
\end{cases}
\end{equation}

We then use non-negative least squares to fit the data twice. 
\begin{enumerate}
\item In the first fit, we use all the normalized templates (ghosts, background, and Gaussian) to measure the amplitude of each prominent ghost.
\item In the second fit, since the saturated source flux is orders of magnitude larger than the ghosts, we mask the region occupied by the Gaussian template and fit only the CBP ghosts and background to ensure that the fits are not skewed due to the source fit.
\end{enumerate}

\subsection{Reflectance Measurements}
\begin{figure}[h!]
    \centering
\includegraphics[width=0.75\textwidth]{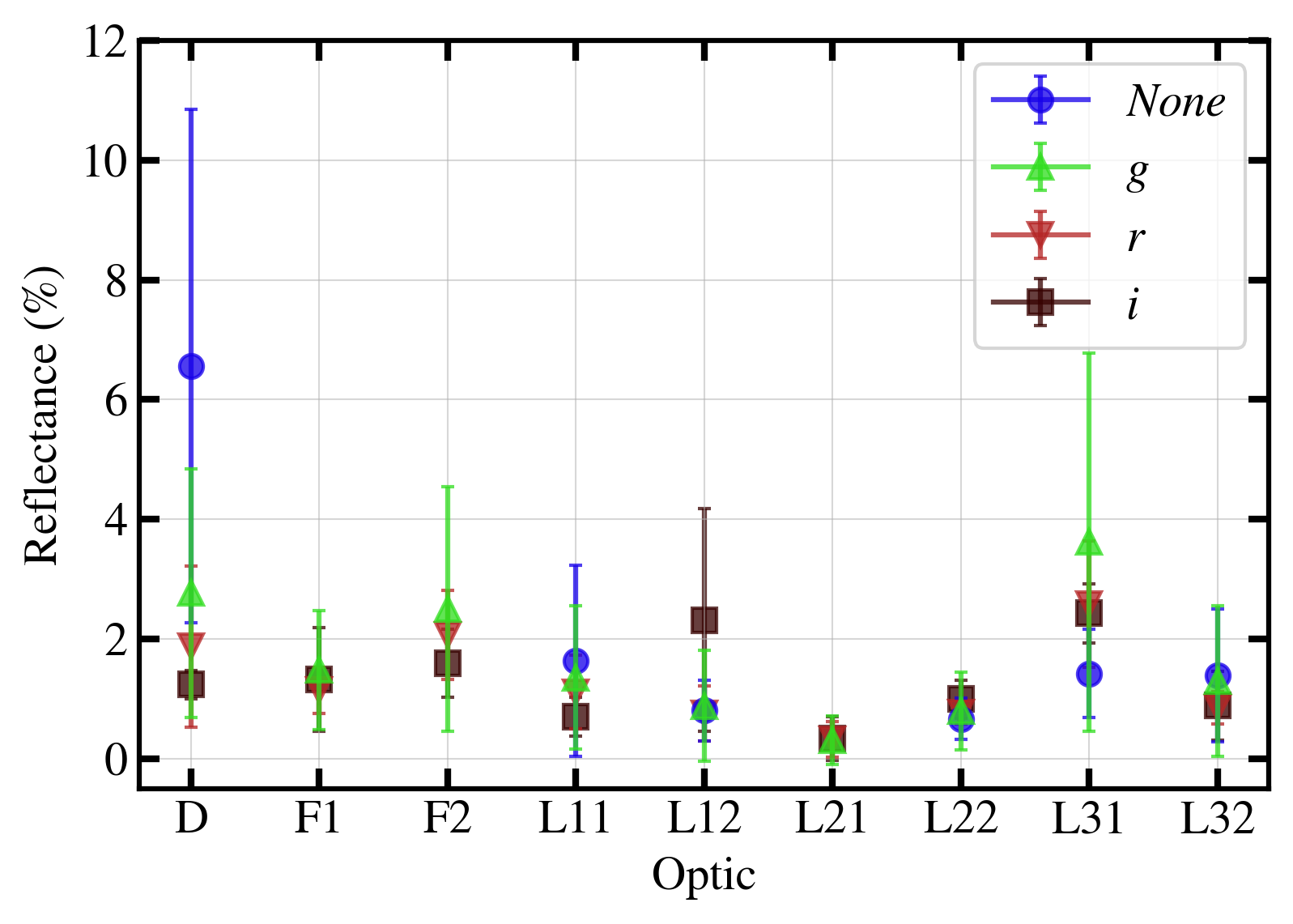}
    \caption{Measured reflectances for each optical surface. The measurements are made at a single wavelength for each optical configuration: no-filter (519\,nm), {\it g} (520\,nm), {\it r} (620\,nm), and {\it i} (760\,nm).} 
    \label{fig:reflectance_vs_optic}
\end{figure}

After measuring the fluxes of the CBP ghosts, we measure the reflectances of the camera optics using Eq.~\ref{eq:reflectance_matrix}. 
Figure~\ref{fig:reflectance_vs_optic} shows the reflectances of each optical element in the no-filter (labelled ``None'') and {\it g}, {\it r}, and {\it i} bands.
Note that measurements for F1, F2 and L21 are not available in the no-filter data (the latter is unavailable as the most prominent ghosts from L21 also require the filter).
More detailed distributions of the reflectance measurements (with both the fit to the source and without are shown in Appendix~\ref{app:reflectance_hists}.

 Table~\ref{tab:syseng_throughputs} shows the simulated reflectance values for the detector and filter (F$_s$ \& D$_s$) produced by the systems engineering simulations. 
 Note that these simulations do not differentiate the two surfaces of the optical elements. 
 The table also contains our measured F1, F2 and D reflectances derived from the masked-source fits and the wavelengths of the laser that we used to produce the CBP ghosts in each band.

\begin{table}[t!]
\centering
\begin{tabular}{cccccccc}
\hline
Band & Wavelength (nm) & F$_s$ (\%) & D$_s$ (\%) & F1 (\%) & F2 (\%) & D (\%) \\
\hline
{\it None} & 519 & N/A & 6.84 & N/A & N/A & 6.56 $\pm$ 4.29 \\
{\it g} & 520 & 2.52 & 6.79 & 1.48 $\pm$ 0.99 & 2.50 $\pm$ 2.04 & 2.77 $\pm$ 2.08\\
{\it r} & 620 & 3.89 & 2.98 & 1.14 $\pm$ 0.38 & 2.07 $\pm$ 0.74 & 1.87 $\pm$ 1.35\\
{\it i} & 760 & 1.54 & 4.15 & 1.32 $\pm$ 0.86 & 1.59 $\pm$ 0.57 & 1.24 $\pm$ 0.24\\
\hline
\end{tabular}
\caption{Systems engineering optical reflectances (F$_s$ \& D$_s$) and our measurements (F1, F2 \& D) for the filter and detector.
Note that the filter is expected to have higher reflectances than the values shown here near the edges of the bandpass.}
\label{tab:syseng_throughputs}
\end{table}

We observe that the source beam is saturated, which could lead to a misestimate of the reflectances due to an inaccurate measurement of the source flux.
Let $F_{\rm src}$ be the true source flux.
% , and the measured source flux be related to the true source flux by a multiplicative factor $\xi$ such that $F_{\rm src} = \xi\,f_{\rm src}$.
We expect that the measured ghost flux is related to the \textit{true} source flux $f_{A, B} = R_AR_BF_{\rm src}$ since the photon reflections are independent of saturation.
Saturation causes the measured source flux $f_{\rm src}$ to underestimate $F_{\rm src}$, inflating the measured ghost-to-source flux ratio $f_{A,B}/f_{\rm src}$ relative to its true value, therefore leading to an overestimation of the reflectances.
Our reflectance measurements can therefore be conservatively taken to be upper bounds.
Assuming that the true source flux is related to the measured source flux by some multiplicative correction factor $\xi$ such that $F_{\rm src} = \xi\,f_{\rm src}$ with $\xi > 1$, the true reflectances would be smaller than the measured values by a factor of $1/\sqrt{\xi}$.
Since the Gaussian fit to the source beam profile constrains $f_{\rm src}$ independently of the saturation, we do not expect $\xi$ to deviate significantly from unity, and the resulting bias in the reflectances is therefore small.
This is further validated by the agreement of our measured 
reflectance values with systems engineering simulations (both are $\sim$2\%).
Finally, we also observe that the measurements of the reflectances with and without masking the source are consistent with each other, leading us to believe that the fit of the source beam is not skewing the measured reflectances.
The large error bar in the measurement of the reflectance of the detector in the no-filter setting is due to the fact that the most significant ghosts produced by the detector are the D-F1 and D-F2 ghosts, which are absent when the filter is not present, thus reducing the constraining power of the model.
 
\section{Conclusions}
We have presented a simulation-based framework for quantifying the impact of optical ghosts on LSSTCam imaging, motivated by the need to assess compliance with the LSST system requirements. 
We combine the Yale Bright Star Catalog, empirically calibrated magnitude transformations, and optical ray-tracing simulations using \texttt{Batoid}, to estimate both the morphology and surface brightness of the dominant ghosts and measure the resulting ghost-impacted area on the focal plane. 
These simulations assumed a 2\% reflectance for each of the optical surfaces, which approximately match  the system engineering simulations.

Applying this framework to $\sim$3,100 visits from the intermittent DRP (w37), we find that the average fraction of the Rubin imaging that is significantly impacted by ghosts is 0.57\% when averaged across all bands.
This value is below the 1\% threshold specified by the LSST system requirement. 
The impacted area exhibits strong band dependence, with the largest contributions occurring in the {\it u} band due to its lower sky background, and minimal impact in the {\it r} and {\it i} bands. 
The impacted area is also heavily dependent on the field being observed, as individual visits containing very bright stars have substantially larger impacted fractions.

We further explored the dependence of ghost-impacted area on stellar magnitude and off-axis position, finding that star brightness is the dominant driver of ghost impact, with only a weak dependence on field angle.
Using these results, we constructed all-sky maps of expected ghost-impacted area based on the Yale Bright Star Catalog. 
Overall, our results demonstrate that optical ghosts are a non-negligible source of contamination for low-surface-brightness science in LSST, particularly in the presence of very bright stars. 

We use the CBP to generate optical ghosts on the focal plane. 
We find good agreement between the morphology of the observed CBP ghosts and those generated by \texttt{Batoid}.
We use \texttt{Batoid} to generate morphological templates for each of the most prominent ghosts and fit the fluxes of the ghosts to derive the reflectances of the optical surfaces.
We find that the reflectances of the optics generally agree with estimates generated by the systems engineering simulations, thus validating the assumption that was made when assessing ghost impact.

\subsection{Acknowledgments}
This material is based upon work supported in part by the National Science Foundation through Cooperative Agreements AST-1258333 and AST-2241526 and Cooperative Support Agreements AST-1202910 and AST-2211468 managed by the Association of Universities for Research in Astronomy (AURA), and the Department of Energy under Contract No.\ DE-AC02-76SF00515 with the SLAC National Accelerator Laboratory managed by Stanford University.
Additional Rubin Observatory funding comes from private donations, grants to universities, and in-kind support from LSST-DA Institutional Members.

Support for AP and ADW was provided by NSF awards AST-2006340 and AST-2307126.
Fermilab is managed by FermiForward Discovery Group, LLC under Contract No.\ 89243024CSC000002 with the U.S. Department of Energy, Office of Science, Office of High Energy Physics. The United States Government retains and the publisher, by accepting the article for publication, acknowledges that the United States Government retains a non-exclusive, paid-up, irrevocable, world-wide license to publish or reproduce the published form of this manuscript, or allow others to do so, for United States Government purposes.

\bibliography{local,lsst,lsst-dm,refs_ads,refs,books}

\bibliographystyle{spiebib} % makes bibtex use spiebib.bst

\newpage
\appendix
\section{Telescope Optic Reflectance Measurements}
\label{app:reflectance_hists}
\begin{figure}[h!]
  \centering

  \begin{subfigure}[t]{0.7\textwidth}
    \centering
    \includegraphics[width=\linewidth]{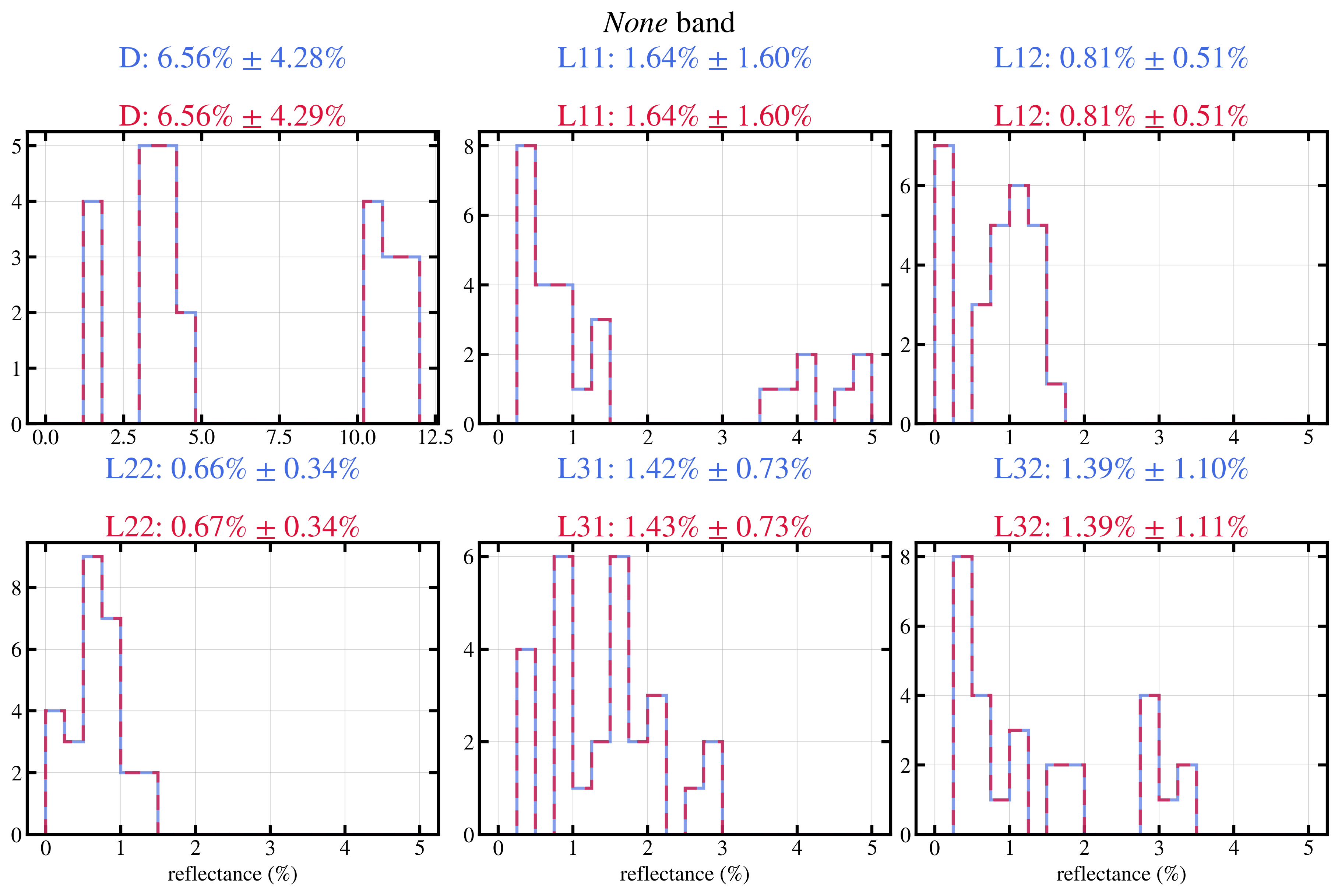}
    \caption{No filter}
    \label{subfig:refl_a}
  \end{subfigure}
  \medskip
  \begin{subfigure}[t]{0.8\textwidth}
    \centering
    \includegraphics[width=\linewidth]{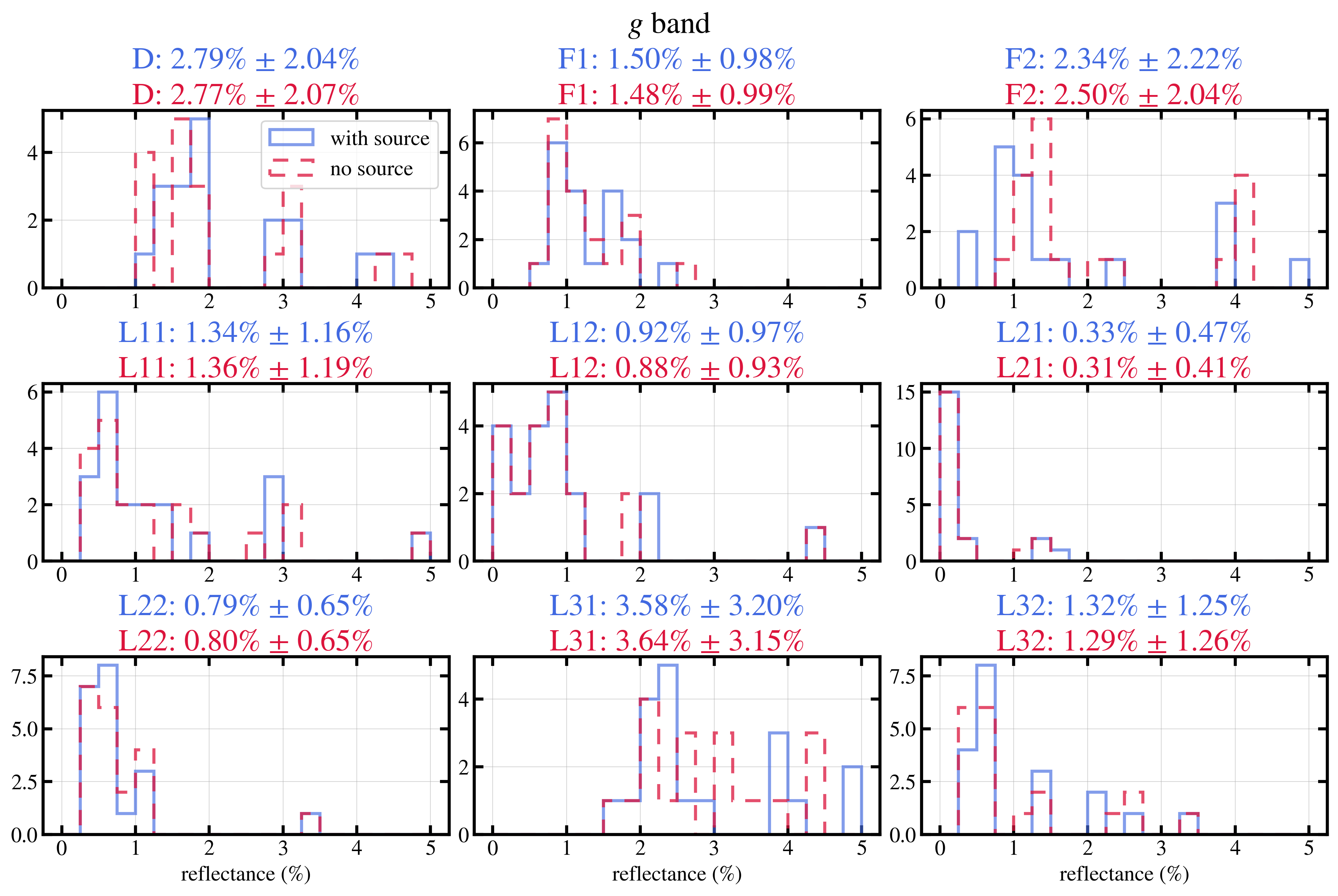}
    \caption{{\it g} band}
    \label{subfig:refl_b}
  \end{subfigure}

  \caption{Distributions of reflectance measurements per visit separated by optic in the no-filter and {\it g} bands.
  The blue measurements are made including a fit to the source and the red measurements are made while masking the source.}
  \label{fig:reflectances_a}
\end{figure}

\begin{figure}[h!]
  \centering

  \begin{subfigure}[t]{0.8\textwidth}
    \centering
    \includegraphics[width=\linewidth]{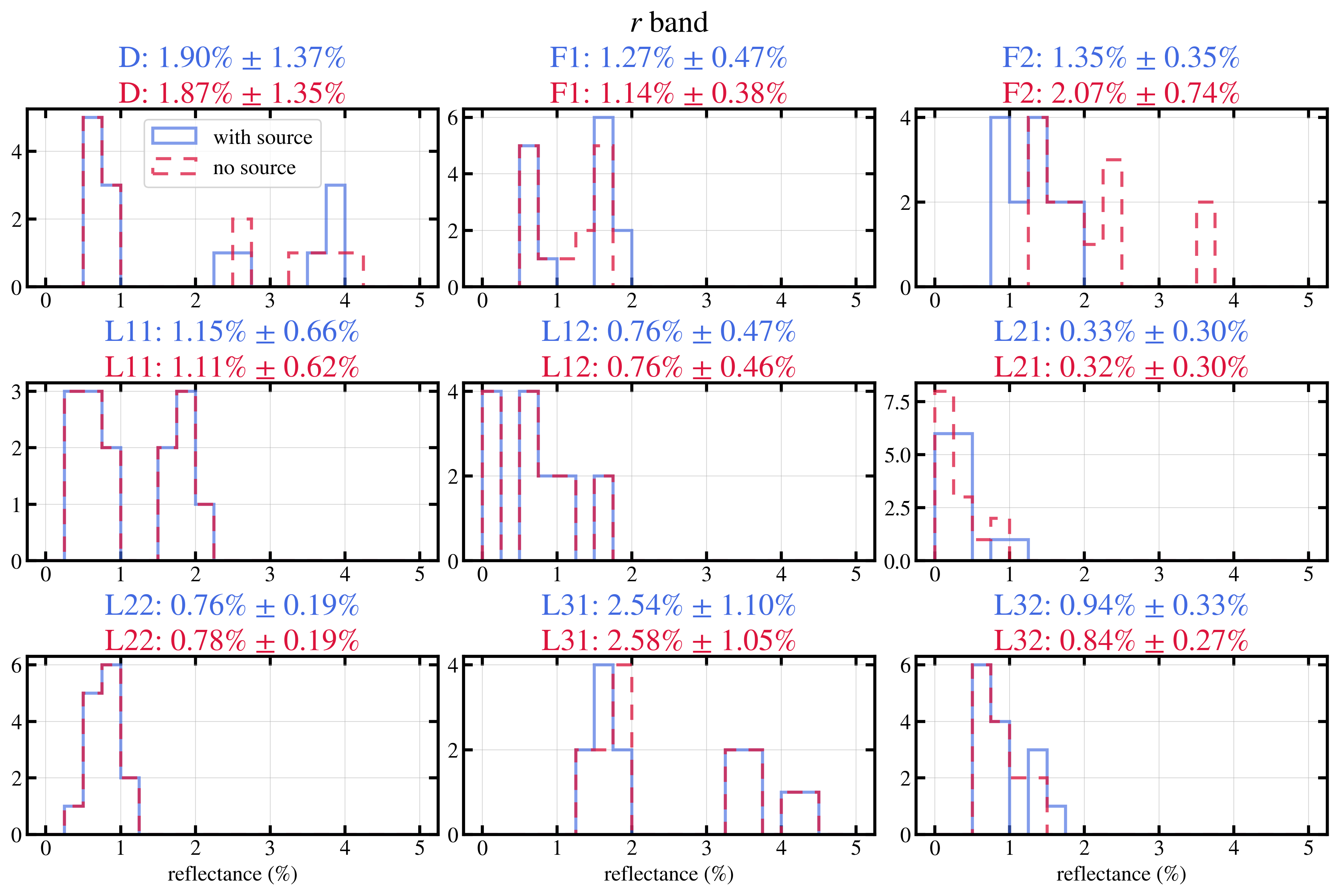}
    \caption{{\it r} band}
    \label{subfig:refl_c}
  \end{subfigure}
 \medskip
  \begin{subfigure}[t]{0.8\textwidth}
    \centering
    \includegraphics[width=\linewidth]{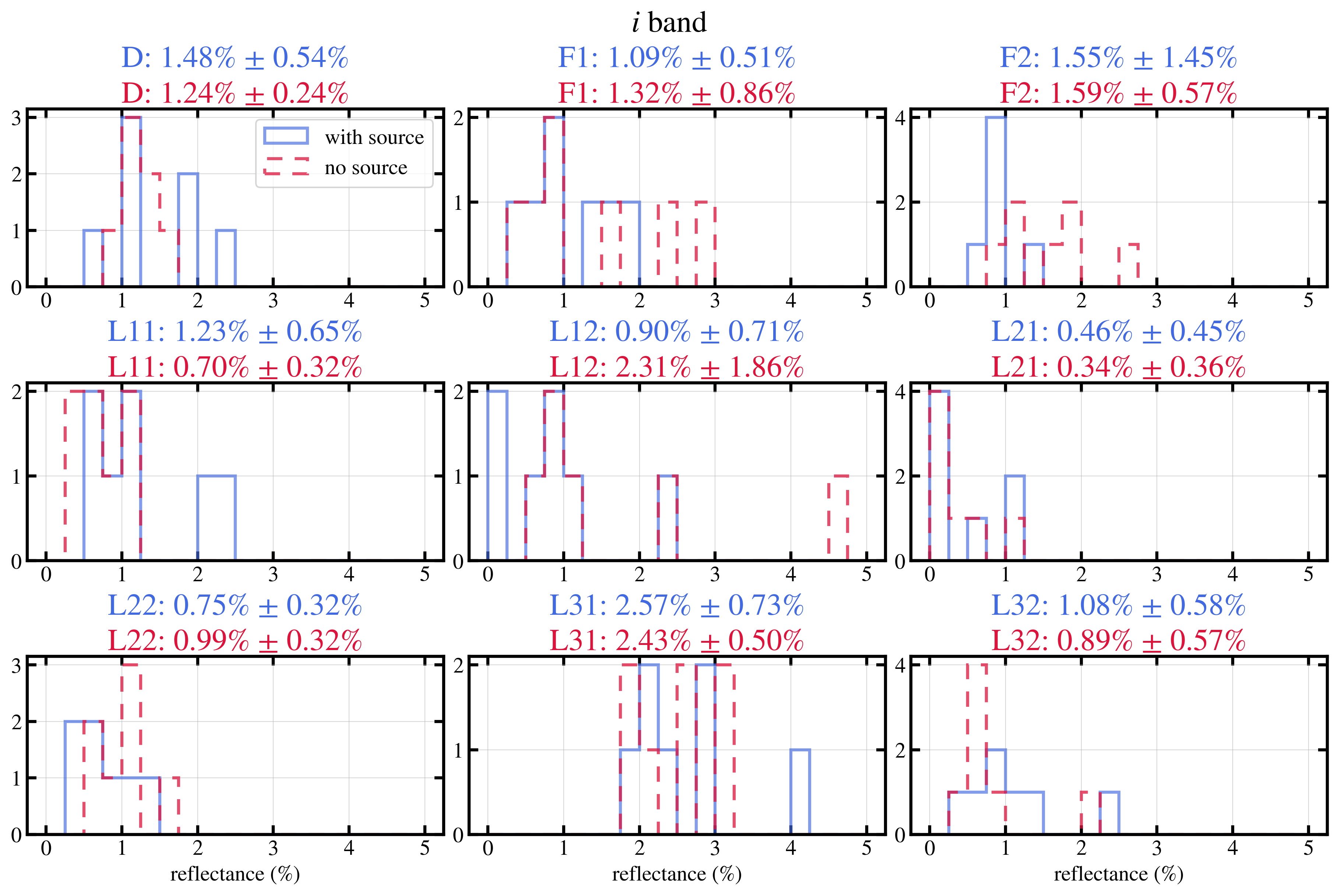}
    \caption{{\it i} band}
    \label{subfig:refl_d}
  \end{subfigure}

  \caption{Distributions of reflectance measurements per visit separated by optic in the {\it r} and {\it i} bands.
  The blue measurements are made including a fit to the source and the red measurements are made while masking the source.}
  \label{fig:reflectances_b}
\end{figure}

\end{document}